%% file: inner_v7.tex
\documentclass[traditabstract]{aa}  
\usepackage{graphicx}
\usepackage{txfonts}
\usepackage{natbib}
\usepackage{longtable}
\usepackage{multirow}

\begin{document}
   \title{Inner rings in disc galaxies: dead or alive}

   \author{S.~Comer\'on
          \inst{1,2}
          }

   \institute{University of Oulu, Astronomy Division, Department of Physics, P.O.~Box 3000, FIN-90014, Finland\\
              \email{seb.comeron@gmail.com}
         \and
             Finnish Centre of Astronomy with ESO (FINCA), University of Turku, V\"ais\"al\"antie 20, FI-21500, Piikki\"o, Finland\\
             }

\authorrunning{Comer\'on, S.}
 
  \abstract
  {In this {\it Letter}, I distinguish ``passive'' inner rings as those with no current star formation as distinct from ``active'' inner rings that have undergone recent star formation. I built a sample of nearby galaxies with inner rings observed in the near- and mid-infrared from the NIRS0S and the S$^4$G surveys. I used archival far-ultraviolet (FUV) and H$\alpha$ imaging of 319 galaxies to diagnose whether their inner rings are passive or active. I found that passive rings are found only in early-type disc galaxies ($-3\leq T\leq2$). In this range of stages, $21\pm3\%$ and $28\pm5\%$ of the rings are passive according to the FUV and H$\alpha$ indicators, respectively. A ring that is passive according to the FUV is always passive according to H$\alpha$, but the reverse is not always true. Ring-lenses form $30-40\%$ of passive rings, which is four times more than the fraction of ring-lenses found in active rings in the stage range $-3\leq T\leq2$. This is consistent with both a resonance and a manifold origin for the rings because both models predict purely stellar rings to be wider than their star-forming counterparts. In the case of resonance rings, the widening may be at least partly due to the dissolution of rings. If most inner rings have a resonance origin, I estimate 200\,Myr to be a lower bound for their dissolution time-scale. This time-scale is of the order of one orbital period at the radius of inner rings.}

   \keywords{Galaxies: evolution -- Galaxies: kinematics and dynamics -- Galaxies: spiral -- Galaxies: statistics}

   \maketitle

\section{Introduction}

\label{introduction}

Gas in disc galaxies is redistributed by angular momentum transfer caused by nonaxisymmetries with a given pattern speed such as bars, ovals, and spiral arms. Some of the gas is collected in orbits near dynamical resonances under the influence of the torques caused by the nonaxisymmetries \citep[for a recent review on barred galaxy dynamics see][]{ATH12B}. Owing to star formation triggered by the high gas density and by gas travelling in intersecting orbits at each side of the resonance, rings and pseudorings are often formed there \citep{SCH81, SCH84}. Historically, this picture has been used to explain resonance rings and pseudorings, but recently an alternative model, called the flux tube manifold theory or manifold theory, postulated that at least some of them are caused by gas and/or stars trapped in tubes of orbits that connect the Lagrangian points at the end of the galaxy bars \citep{RO06, RO07, ATH09A, ATH09B, ATH10, ATH12A}. Another alternative ring formation mechanism is that developed by \citet{KIM12} for nuclear rings (not studied here), where rings are formed due to the centrifugal barrier encountered by gas migrating to inner regions of the galaxy. In this {\it Letter}, the word rings is used to refer to both rings and pseudorings.

In the classical resonance theory and also in the manifold theory, inner rings in barred galaxies have a diameter slightly larger than the bar length. In the classic resonance theory they are associated to the ultraharmonic 4:1 resonance. In this {\it Letter}, broad features intermediate between inner lenses and inner rings, called ring-lenses, are accounted for alongside inner rings. Simulations by \citet{RAU00} showed that inner rings, although changing in shape and size, are long-lived, provided there is a gas inflow to feed them. In the manifold theory they are also long-lived as long as the galaxy potential does not evolve, or does so slowly \citep{ATH10}.

``Passive'' inner rings are here those not found to host star formation. In the context of nuclear rings, passive rings have also been called fossil rings \citep{ER01}. As opposed to passive features, ``active'' rings are those with indications of recent star formation. An example of a passive inner ring in the literature is that in NGC~7702 \citep{BU91}. Here, I study inner rings identified in two infrared surveys of nearby galaxies, namely the Spitzer Survey of Stellar Structure in Galaxies \citep[S$^4$G;][]{SHETH10} and the Near-InfraRed atlas of S0-Sa galaxies \citep[NIRS0S;][]{LAU11}. The S$^4$G is a survey of 2352 nearby galaxies in $3.6$ and $4.5\mu$m using the {\it Spitzer} Space Telescope and NIRS0S is a $K_{\rm s}$-band ground-based survey of 206 early-type galaxies. To investigate whether rings are passive or active, I used archival images in one band and one line that trace recent star formation, namely the far-ultraviolet (FUV) and H$\alpha$.

This {\it Letter} is structured as follows. In Section~\ref{data}, I present the sample, the data, and the image processing. Then, I present the results in Section~\ref{results} and discuss them in Section~\ref{discussion}. The conclusions are summarized in Section~\ref{conclusions}.

\section{Data selection and processing}

\label{data}

\begin{figure}
\begin{center}
  \includegraphics[width=0.45\textwidth]{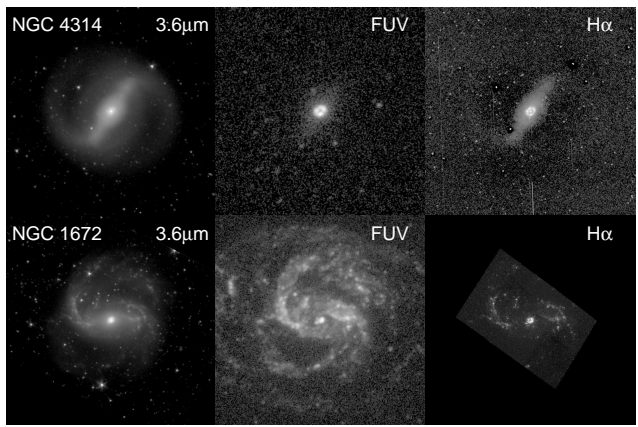}
  \caption{\label{examples} NGC~4314 (top row) is an (R$_1^{\prime}$)SB(rl,nr)a galaxy \citep{BU13} with a passive inner ring-lens. NGC~1672 (bottom row) is an (R$^{\prime}$)SA${\underline{\rm B}}$(rs,nr)b galaxy \citep{BU13} whose inner pseudoring is still forming stars. In the first case the inner feature does not emit in the FUV and H$\alpha$. There is a significant amount of emission in these bands for the active inner feature. The prominent ring feature in NGC~4314 is a nuclear ring. The $3.6\,\mu{\rm m}$ images are from the S$^4$G, the FUV images come from GALEX's NGS survey, and the H$\alpha$ images come from \citet{KNA04} and the HLA.}
\end{center}
\end{figure}

I mainly based my sample of galaxies with inner rings on the classification of S$^4$G galaxies made by \citet{BU13} and statistically studied in the Atlas of Resonance (pseudo)Rings As Known In the S$^4$G \citep[ARRAKIS;][]{CO13}. Since the S$^4$G sample is biased against galaxies with a small gas fraction, which are mostly elliptical and S0 galaxies, I also included NIRS0S galaxies with inner rings that matched the S$^4$G selection criteria, namely galactic latitude $|b|>30^{\rm o}$, radial velocity $v_{\rm r}<4000\,{\rm km\,s^{-1}}$, angular diameter $D_{25}>1\arcmin$, and integrated blue magnitude $m_{\rm B}<15.5$\,mag \citep[data obtained from HyperLeda;][]{PA03}. S$^4$G and NIRS0S data can be mixed safely because the detection of inner rings in the S$^4$G matches that in NIRS0S very well (Section~5.10 in ARRAKIS). I also included NGC~2950, an S0 non-S$^4$G galaxy appearing in the same frame as a genuine S$^4$G galaxy which also fulfils the selection criteria.

To avoid dust absorption, ring foreshortening, and poor angular resolution problems, I additionally constrained the sample by only selecting disc galaxies (Hubble stage $-3\leq T\leq9$) with an ellipticity lower than $\epsilon_{\rm d}=0.5$ according to the data of the Pipeline~4 of S$^4$G \citep{SA13} and with inner rings with a radius larger than 10\arcsec\ according to ARRAKIS or NIRS0S. The total number of galaxies fulfilling these conditions is 357.

Two indicators were used to search for recent star formation: the far-ultraviolet continuum and H$\alpha$-line emission.

The ultraviolet continuum traces star formation that has occurred in the past 100\,Myr \citep{KEN98}. To study inner rings in that wavelength, I downloaded the deepest available FUV-band image in the GALEX GR6/7 Data Release\footnote{http://galex.stsci.edu/GR6/} for each galaxy. Such images were available for 319 out of the 357 galaxies initially included in the sample. These 319 galaxies are the final sample I worked with. For 160 galaxies, the FUV images belong to the GALEX All-Sky Imaging Survey (AIS), which consists of $\sim100$\,s exposures and can detect point sources down to $\mu_{\rm AB}\sim20\,{\rm mag\,arcsec^{-2}}$. The other galaxies were imaged in deeper GALEX surveys and were in general exposed for 1000\,s or more.

The H$\alpha$ emission traces star formation that has occurred in the past 20\,Myr \citep{KEN98}. H$\alpha$ continuum-subtracted images used here come from three sources:
\begin{itemize}
 \item Images processed for the Atlas of Images of NUclear Rings \citep[AINUR;][]{CO10}. The images in AINUR come mostly from the $Hubble$ Space Telescope (HST) Archive\footnote{http://archive.stsci.edu/hst/search.php}.
 \item HST images not processed for AINUR. In these cases, H$\alpha$ narrow-band images and red continuum images were downloaded from the {\it Hubble} Legacy Archive\footnote{http://hla.stsci.edu/hlaview.html} (HLA) and were used to produce a continuum-subtracted image using the technique described in \citet{KNA04, KNA06}.
 \item Continuum-subtracted images in the NASA/IPAC Extragalactic Database (NED)\footnote{http://ned.ipac.caltech.edu/}.
\end{itemize}

H$\alpha$ images were available for 139 out of the 319 sample galaxies.

For each sample galaxy I verified in the FUV and H$\alpha$ continuum-subtracted images whether the rings detected in S$^4$G or NIRS0S images were visible. A detection was considered to be positive if at least a segment of the ring was seen. In some doubtful cases with shallow AIS FUV images, this was only possible after smoothing the image with a Gaussian kernel with a 3-pixel (4.5\arcsec) radius. Positive detections are labelled as ``A'' for ``active'' and negative detections are labelled as ``P'' for ``passive'' in Table~\ref{table}. In doubtful cases, a ``?'' sign is added to the detection status. From now on, I refer to positive and negative detections as active and passive rings, respectively. This definition means that some rings may be passive in one of the studied indicators but active in the other. Examples of galaxies with passive and active inner rings are shown in Figure~\ref{examples}. A total of 329 inner rings in the 319 sample galaxies were analysed.

\section{Results}

\label{results}

\begin{figure}
\begin{center}
  \includegraphics[width=0.45\textwidth]{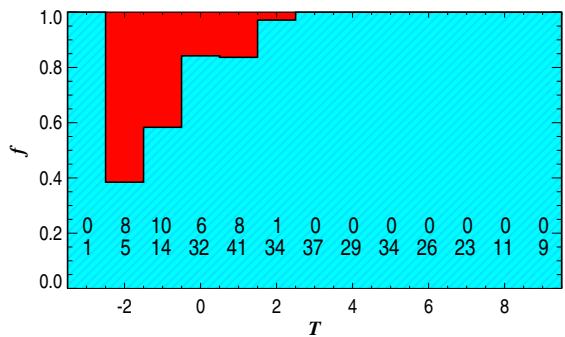}
  \caption{\label{fir} Fraction of active and passive inner rings (blue hatched and red plain surfaces, respectively) according to the FUV indicator as a function of the galaxy stage for galaxies in the sample with available GALEX imaging. The bottom row of numbers indicate the number of active rings for a given stage and the top row indicates the number of those that are passive.}
\end{center}
\end{figure}

\begin{figure}
\begin{center}
  \includegraphics[width=0.45\textwidth]{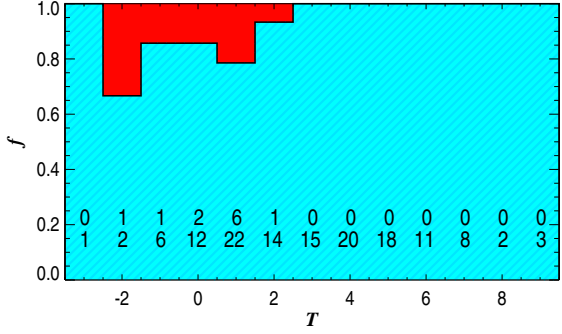}
  \caption{\label{haandfir} Fraction of active and passive inner rings according to the FUV indicator as in Figure~\ref{fir}, but now for inner rings in galaxies that have both FUV and H$\alpha$ imaging available.}
\end{center}
\end{figure}

\begin{figure}
\begin{center}
  \includegraphics[width=0.45\textwidth]{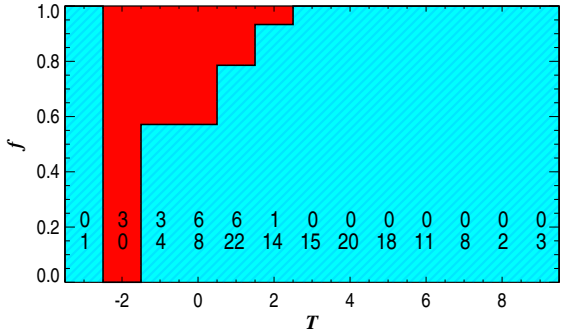}
  \caption{\label{ha} As Figure~\ref{fir}, but now using H$\alpha$ emission as an indicator on whether inner rings are passive or active.}
\end{center}
\end{figure}

\begin{table}
 \caption{Fraction of passive rings according to different star formation indicators and for different Hubble stage ranges}
 \label{stats1}
 \centering
 \begin{tabular}{l c c c c}
  \hline
  Indicator & All types       & S0              & Early sp.     & Late sp.      \\
            & $-3\leq T\leq9$ & $-3\leq T\leq0$ & $1\leq T\leq2$& $3\leq T\leq9$\\
  \hline
  \hline
  FUV       & $10\pm2\%$      & $32\pm5\%$      & $11\pm3\%$    & $0\%$         \\
  Deep FUV  & $13\pm3\%$      & $36\pm8\%$      & $15\pm5\%$    & $0\%$         \\
FUV (H$\alpha$)& $8\pm2\%$    & $16\pm7\%$      & $16\pm6\%$    & $0\%$         \\
  H$\alpha$ & $13\pm3\%$      & $48\pm10\%$     & $16\pm6\%$    & $0\%$         \\
  \hline
 \end{tabular}
   \tablefoot{Deep FUV stands for GALEX FUV imaging excluding images from the AIS survey. FUV (H$\alpha$) stands for the FUV statistics assembled from galaxies with available H$\alpha$ images.}
\end{table}

\begin{table*}
 \caption{Fraction of passive and active inner features with a given morphology as classified in \citet{BU13} for different star formation indicators in galaxies with $-3\leq T\leq2$.}
 \label{stats2}
 \centering
 \begin{tabular}{l | c c | c c | c c}
  \hline
  Indicator & \multicolumn{2}{c|}{r${\underline{\rm l}}$, rl, ${\underline{\rm r}}$l, r$^{\prime}{\underline{\rm l}}$, r$^{\prime}$l, ${\underline{\rm r}}^{\prime}$l} & \multicolumn{2}{c|}{r} & \multicolumn{2}{c}{r${\underline{\rm s}}$, rs, ${\underline{\rm r}}$s}     \\
               &    P     &    A    &   P      &   A      &   P      &    A     \\
  \hline
  \hline
     FUV       &$39\pm9\%$&$9\pm3\%$&$33\pm7\%$&$35\pm4\%$&$27\pm8\%$&$56\pm4\%$\\
FUV (H$\alpha$)&$36\pm15\%$&$9\pm4\%$&$36\pm15\%$&$32\pm6\%$&$27\pm13\%$&$60\pm6\%$\\
     H$\alpha$ &$32\pm11\%$&$6\pm3\%$&$32\pm11\%$&$33\pm7\%$&$37\pm11\%$&$61\pm7\%$\\
  \hline
 \end{tabular}
 \tablefoot{P and A stand for passive and active rings, respectively. FUV (H$\alpha$) stands for the FUV statistics assembled from galaxies with available H$\alpha$ images.}
\end{table*}

Out of 329 inner rings, 33 were found not to have FUV emission (Figure~\ref{fir}). All of them are in the stage range $-3\leq T\leq2$. The fraction of passive inner rings in this range of stages according to the FUV star formation indicator is $21\pm3\%$ with the error bar calculated using binomial statistics. Rings hosted in early-type galaxies are more likely to be passive than those in later types (Table~\ref{stats1}). I verified whether non-detections may be partly caused by the use of shallow AIS images by recalculating the statistics with deep GALEX images only (Table~\ref{stats1}) and found that the results based on those two samples are compatible within the error bars, which indicates that rings with recent star formation can be detected even in relatively shallow FUV images.

Because H$\alpha$ imaging is only available for a part of the sample, I reproduced the plot for the FUV inner ring emission in Figure~\ref{fir} by using only FUV data that correspond to galaxies for which H$\alpha$ is available (Figure~\ref{haandfir} and Table~\ref{stats1}). This is used below to compare the fraction of passive inner rings according to the FUV and H$\alpha$ indicators.

Figure~\ref{ha} and Table~\ref{stats1} show the fraction of passive inner rings according to the H$\alpha$ indicator. This fraction is equal to or larger than that of passive rings according to the FUV for all stages ($28\pm5\%$ of passive rings in the range $-3\leq T\leq2$ where all passive rings are found). This is because none of the rings lacking FUV emission have H$\alpha$ emission, whereas the reverse is not always true.

While in the range of stages $-3\leq T\leq 2$ the fraction of active inner features classified as ring-lenses is one in ten or less, $30-40\%$ of the passive inner features are ring-lenses (Table~\ref{stats2}). Regarding inner features that are not classified as ring-lenses, inner closed rings are equally frequent among the passive and the active rings ($\sim30-40\%$), but pseudorings are less frequent among passive features than among active ones ($\sim30\%$ vs $\sim60\%$).

The passive or active status of a ring does not depend on the family (bar properties) of the host galaxy. Unbarred galaxies (SA) account for $\sim30\%$ of host galaxies for both passive and active rings in the stage range $-3\leq T\leq2$.

\section{Discussion}

\label{discussion}

As seen in Section~\ref{introduction}, two mechanisms have been proposed for the formation of inner rings, namely the resonance and the manifold ones. I consider the resonance mechanism first.

If inner rings are the consequence of the star formation in gas gathered in orbits near resonances, one may expect that once the gas is exhausted, the ring will fade-out and disappear because of two factors. First, bright stars with a low mass-to-light ratio will die after several Myr. Second, radial migration will cause the ring to widen. Both effects would lower the surface brightness of the ring and will tend to make it indistinguishable from the stellar background of the galaxy. Of course, active rings may also have populations of old stars widened by radial migration, but they are likely to be outshone by the regions with recent star formation and thus would appear very sharp.

One piece of evidence that might indicate that rings become wider as they die is that as seen in Table~\ref{stats2}, ring-lenses are roughly four times more frequent among passive features than among active features. It is unclear, however, whether the full width of ring-lenses can be explained by the radial migration of stars in passive rings. Alternatively, ring-lenses may form as a response of the old stellar population to the bar potential.

Under the assumption that the resonance scenario applies, an estimate of the dissolution time-scale of rings can be made from the data presented here and by knowing that H$\alpha$ emission outlasts star formation by $\sim20$\,Myr and FUV emission outlasts star formation by $\sim100$\,Myr \citep{KEN98}. The subsample of galaxies with both H$\alpha$ and FUV imaging includes 19 rings without H$\alpha$ emission. Eleven of those rings have no FUV emission. This means that eight rings stopped forming stars between 20 and 100\,Myr ago, and the remainder stopped forming stars longer than 100\,Myr ago. Assuming that the fraction of dissolving inner rings has been roughly constant for the past few hundreds of Myr, one can deduce that the ring dissolution time-scale is $\sim200$\,Myr. This is a time of the order of an orbital period at the radius of inner rings.

However, this $\sim200$\,Myr estimated dissolution time-scale is a lower limit to the true dissolution time-scale. First, rings may form stars intermittently in recurrent episodes. It is therefore reasonable to assume that some of the passive rings may actually be reactivated at some point by some gas inflow. Such periodic activity has been reported in nuclear rings \citep{ALL06, SAR07}. Episodic star formation seems more likely in rings that have stopped forming stars more recently (those without H$\alpha$ emission but with FUV emission), hence the dissolution time-scale underestimation. Second, H$\alpha$ surveys may be biased against galaxies with little or no H$\alpha$ emission. This would bias the surveys against galaxies with passive rings, and especially against those that cannot be reactivated, because if the galaxy still has some gas reserve that can be transferred to the ring, some residual star formation may remain elsewhere in the galaxy. As a consequence, the fraction of inner rings without either FUV or H$\alpha$ emission might be underestimated.

In the framework of the manifold theory, passive rings are not necessarily dissolving. Indeed, manifolds can trap both stars and gas, and for galaxies with little or no gas, purely stellar rings are expected. However, it is still natural to expect passive rings to be wider. Indeed, stars can easily occupy the whole manifold phase space, but gas collisions would cause it (and also the younger generations of stars) to fill a smaller space and thus make the rings appear thinner \citep{ATH09A}. Whether this effect is enough to explain the full difference in width between regular rings and ring-lenses is not yet explored.

A large fraction of inner rings with a manifold origin could significantly change the dissolution time-scale estimated before. At the moment, no estimate is available on the fraction of inner rings caused by manifolds. Their existence is nearly certain, however, because the characteristic morphology of all types of outer rings, as well as the statistics of the shapes and sizes of both inner and outer rings in nearby galaxies, can be explained by the manifold theory \citep{ATH09A, ATH10}. On the other hand, a set of $\sim20$ N-body simulations with a fixed potential shows that at least in some cases only a minority of ring particles are trapped in manifolds (P.~Rautiainen, private communication). This, however, is not the case for the fully self-consistent simulations of \citet{ATH12A}. Additional study, both observational and numerical, is required to reveal whether manifolds can be easily populated and therefore are a widespread mechanism for shaping galaxy morphology.

In either the resonance or manifold frameworks the lack of passive rings in galaxies with $T\ge3$ is naturally explained because in both cases, gas is available in these late-type galaxies to populate the orbits near resonances and/or the manifold orbits.

\section{Conclusions}

I used two indicators of recent star formation to check whether inner rings and pseudorings in a set of 319 nearby disc galaxies are passive (without signs of recent star formation) or active (with signs of recent star formation).

I showed that passive rings are only found in galaxies with stages $-3\leq T\leq2$. In that range of stages, $21\pm3\%$ and $28\pm5\%$ of rings are passive according to the FUV and H$\alpha$ indicators, respectively. When a ring is passive in the FUV, it is also passive in H$\alpha$, but the reverse is not always true. I found that $30-40\%$ of passive inner rings are classified as ring-lenses in \citet{BU13}. On the other hand, only $\sim10\%$ of active inner rings in the stage range $-3\leq T\leq2$ are ring-lenses. Although passive rings in both resonance and manifold theories are expected to be wider than their active counterparts, it is still unclear whether these two theories can account for the full transformation of regular inner rings into wide inner ring-lenses.

I estimate that if most inner rings have a resonance origin, a lower boundary for their dissolution time-scale is 200\,Myr. This time-scale is of the order of one orbital period at the radius of inner rings.

\label{conclusions}

\begin{acknowledgements}
The author thanks the referee, S.~Ryder for useful comments. He thanks L.~C.~Ho, who gave the inspiration for this {\it Letter} and P.~Rautiainen, H.~Salo, J.~H.~Knapen and E.~Athanassoula for useful discussions. GALEX (Galaxy Evolution Explorer) is a NASA Small Explorer, launched in April 2003. We gratefully acknowledge NASA's support for construction, operation, and science analysis for the GALEX mission. This research has made use of the NASA/IPAC Extragalactic Database (NED) which is operated by the Jet Propulsion Laboratory, California Institute of Technology, under contract with the National Aeronautics and Space Administration. Based on observations made with the NASA/ESA Hubble Space Telescope, and obtained from the Hubble Legacy Archive, which is a collaboration between the Space Telescope Science Institute (STScI/NASA), the Space Telescope European Coordinating Facility (ST-ECF/ESA) and the Canadian Astronomy Data Centre (CADC/NRC/CSA). 
\end{acknowledgements}

\bibliographystyle{aa}
\bibliography{inner}

\Online
\appendix
\onecolumn
\section{Properties of the inner rings in the sample}
\begin{longtable}{l c c c c c c}
\caption{\label{table}Properties of the inner rings in the sample}\\
\hline
ID & Family & $T$ & Kind & FUV & Survey & H$\alpha$\\
(1)&(2)&(3)&(4)&(5)&(6)&(7)\\
\hline\hline
\endfirsthead
\caption{continued.}\\
\hline
ID & Family& $T$ & Kind & FUV & Survey & H$\alpha$\\
(1)&(2)&(3)&(4)&(5)&(6)&(7)\\
\hline\hline
\endhead
\hline
\endfoot
\hline
\endlastfoot
\input{table.tex}
\end{longtable}
ID (Column~1) refers to the galaxy name, family (Column~2) to its bar classification and and $T$ (Column~3) to its stage \citep[from][and NIRSOS]{BU13}. Kind (Column~4) indicates the ring classification by \citet{BU13} and NIRS0S. FUV (Column~5) indicates whether a given ring emits in the ultraviolet continuum (``A'') or not (``P''). The Survey column (Column~6) indicates to which GALEX survey the FUV images used here belong. H$\alpha$ (Column~7) indicates whether a given ring is seen in continuum-subtracted H$\alpha$ images (``A''), or not (``P''). Uncertain detection statuses are indicated by ``?'' in Columns~5 and 7.
\end{document}

%% file: table.tex
ESO~013-016&SB&6&rs&A&AIS&$-$\\
ESO~202-041&SB&9&rs&A&AIS&$-$\\
ESO~404-012&SAB&3&${\underline{\rm r}}$s&A&AIS&$-$\\
ESO~407-014&SA&2&r&A&MIS&$-$\\
ESO~420-009&S${\underline{\rm A}}$B&5&rs&A&AIS&$-$\\
ESO~422-005&SB&8&r${\underline{\rm s}}$&A&AIS&$-$\\
ESO~440-011&SB&6&rs&A&NGS&$-$\\
ESO~443-069&SB&8&rs&A&AIS&$-$\\
ESO~479-004&SB&7&r${\underline{\rm s}}$&A&GII&$-$\\
ESO~482-035&SB&3&rs&A&AIS&$-$\\
ESO~508-007&SA${\underline{\rm B}}$&7&rs&A?&AIS&$-$\\
ESO~510-059&SB&5&r${\underline{\rm s}}$&A&AIS&$-$\\
ESO~532-022&SB&7&r${\underline{\rm s}}$&A&AIS&$-$\\
ESO~547-005&SAB&9&rs&A&MIS&$-$\\
ESO~548-005&SAB&8&rs&A&AIS&$-$\\
ESO~572-018&SA${\underline{\rm B}}$&3&rs&A&AIS&$-$\\
ESO~576-032&SB&5&rs&A&AIS&$-$\\
ESO~602-030&SB&7&r${\underline{\rm s}}$&A&AIS&$-$\\
IC~0749&SB&6&rs&A&AIS&A\\
IC~1014&SB&9&r&A&AIS&$-$\\
IC~1067&SB&3&r&A&GII&$-$\\
IC~1265&SA&0&r&A?&AIS&$-$\\
IC~1438&SAB&0&r$^{\prime}$l&A&AIS&$-$\\
IC~1954&SB&6&r${\underline{\rm s}}$&A&GII&$-$\\
IC~2969&SA&7&r&A&AIS&$-$\\
IC~3102&SAB&0&${\underline{\rm r}}$s&A&GII&$-$\\
IC~3267&SA&1&rs&A&GII&$-$\\
IC~4214&S${\underline{\rm A}}$B&0&r$^{\prime}$l&A&AIS&$-$\\
IC~4237&SB&3&r&A&AIS&$-$\\
IC~5240&SB&0&r&A&AIS&A\\
IC~5267&SA&0&r&A&AIS&$-$\\
NGC~0150&SAB&2&r${\underline{\rm s}}$&A&AIS&$-$\\
NGC~0210&SAB&2&r$^{\prime}$l&A&GII&A\\
NGC~0255&SB&6&rs&A&AIS&$-$\\
NGC~0289&SA${\underline{\rm B}}$&2&rs,${\underline{\rm r}}$s&A,A&MIS&$-$\\
NGC~0470&S${\underline{\rm A}}$B&2&r${\underline{\rm s}}$&A&MIS&A\\
NGC~0473&SA&-1&r&A&AIS&A\\
NGC~0488&SA&1&rl&P&MIS&P\\
NGC~0600&SB&7&r${\underline{\rm s}}$&A&AIS&$-$\\
NGC~0613&SB&3&${\underline{\rm r}}$s&A&AIS&A\\
NGC~0658&SA&4&r${\underline{\rm s}}$&A&MIS&A\\
NGC~0691&SA&2&rs,r&A,A&AIS&A,A\\
NGC~0701&SB&7&${\underline{\rm r}}$s&A&MIS&$-$\\
NGC~0718&SAB&1&rs&P?&AIS&P\\
NGC~0864&SA${\underline{\rm B}}$&4&r${\underline{\rm s}}$&A&MIS&A\\
NGC~0908&SA&3&rs&A&GII&$-$\\
NGC~0936&SB&-1&${\underline{\rm r}}$s&P&GII&$-$\\
NGC~0941&S${\underline{\rm A}}$B&5&r&A&GII&A\\
NGC~0986&SB&2&rs&A&NGS&A\\
NGC~1022&SAB&0&${\underline{\rm r}}$s&A&NGS&P\\
NGC~1073&SB&5&rs&A&GII&A\\
NGC~1079&S${\underline{\rm A}}$B&-1&${\underline{\rm r}}$s&A&MIS&$-$\\
NGC~1087&SB&7&r${\underline{\rm s}}$&A&MIS&A\\
NGC~1097&SB&3&rs&A&NGS&A\\
NGC~1179&SA${\underline{\rm B}}$&6&r${\underline{\rm s}}$&A&MIS&A\\
NGC~1187&SA${\underline{\rm B}}$&4&rs&A&GII&A\\
NGC~1201&SAB&-2&r$^{\prime}$l&P?&AIS&$-$\\
NGC~1232&SAB&5&rs&A&AIS&$-$\\
NGC~1297&SA&-2&rl&P?&MIS&$-$\\
NGC~1310&SB&6&rs&A&NGS&$-$\\
NGC~1317&SAB&0&r$^{\prime}$l&A&NGS&P\\
NGC~1326&SAB&-1&r&A&NGS&P\\
NGC~1350&SAB&0&r&A&NGS&A\\
NGC~1357&SA&0&${\underline{\rm r}}$s&A&AIS&$-$\\
NGC~1365&SB&4&r${\underline{\rm s}}$&A&NGS&$-$\\
NGC~1367&SAB&0&rs&A&GII&P\\
NGC~1385&SB&8&r${\underline{\rm s}}$&A&NGS&$-$\\
NGC~1398&SB&1&${\underline{\rm r}}$s&A&GII&A\\
NGC~1433&SB&1&r&A&AIS&A\\
NGC~1440&SB&-2&${\underline{\rm r}}$s&P?&AIS&$-$\\
NGC~1452&SB&0&r&A?&AIS&$-$\\
NGC~1493&SB&5&rs&A&GII&A\\
NGC~1512&SB&1&r&A&NGS&A\\
NGC~1553&SA&-1&rl&P&NGS&$-$\\
NGC~1566&SAB&3&r$^{\prime}$l&A&NGS&A\\
NGC~1640&SB&1&r&A&AIS&$-$\\
NGC~1672&SA${\underline{\rm B}}$&3&rs&A&NGS&A\\
NGC~2460&S${\underline{\rm A}}$B&1&rs&A&AIS&A\\
NGC~2523&SB&2&r&A&AIS&$-$\\
NGC~2604&SB&5&r${\underline{\rm s}}$&A&MIS&A\\
NGC~2608&SA${\underline{\rm B}}$&3&r${\underline{\rm s}}$&A&MIS&A\\
NGC~2633&SAB&3&rs&A&AIS&A\\
NGC~2681&S${\underline{\rm A}}$B&0&${\underline{\rm r}}$s&A&NGS&P?\\
NGC~2775&SA&-1&${\underline{\rm r}}$s&A&MIS&A\\
NGC~2780&SB&1&rs&A&AIS&$-$\\
NGC~2782&SA&1&rs&A&NGS&$-$\\
NGC~2787&SB&-2&r&A&AIS&P\\
NGC~2805&S${\underline{\rm A}}$B&5&r${\underline{\rm s}}$&A&AIS&A\\
NGC~2859&SAB&-1&rl&A?&GII&P\\
NGC~2906&SA&3&rs&A&MIS&$-$\\
NGC~2950&SA${\underline{\rm B}}$&-1&r${\underline{\rm l}}$&P&AIS&P\\
NGC~2962&S${\underline{\rm A}}$B&-1&rl&P&MIS&$-$\\
NGC~2964&S${\underline{\rm A}}$B&3&r${\underline{\rm s}}$&A&NGS&$-$\\
NGC~2966&SA${\underline{\rm B}}$&1&r$^{\prime}$l&A&MIS&$-$\\
NGC~2967&SAB&5&rs&A&MIS&$-$\\
NGC~2968&SB&-1&r${\underline{\rm s}}$&P&NGS&$-$\\
NGC~2974&SA&0&r&A&GII&$-$\\
NGC~3031&SA&1&rs,r&A,P&GII&A,P\\
NGC~3032&SA&-2&rs,r&P&GII&$-$\\
NGC~3061&SAB&3&${\underline{\rm r}}$s&A&AIS&$-$\\
NGC~3147&S${\underline{\rm A}}$B&3&r${\underline{\rm s}}$&A&NGS&$-$\\
NGC~3166&SB&-1&rl&A&GII&$-$\\
NGC~3184&SA&4&r${\underline{\rm s}}$&A&AIS&A\\
NGC~3185&SAB&1&${\underline{\rm r}}$s&A&NGS&A\\
NGC~3245&SAB&-2&rs&A?&GII&P\\
NGC~3344&SAB&4&r&A&NGS&A\\
NGC~3346&SB&6&rs&A&AIS&$-$\\
NGC~3351&SB&1&r&A&NGS&A\\
NGC~3359&SB&7&rs&A&NGS&A\\
NGC~3368&SAB&-1&${\underline{\rm r}}$s&A&NGS&A\\
NGC~3380&SA${\underline{\rm B}}$&0&${\underline{\rm r}}$s&A&AIS&$-$\\
NGC~3381&SB&8&r${\underline{\rm s}}$&A&AIS&$-$\\
NGC~3455&SA&5&rs&A&GII&$-$\\
NGC~3485&SA${\underline{\rm B}}$&3&${\underline{\rm r}}$s&A&GII&A\\
NGC~3486&SAB&5&r&A&GII&A\\
NGC~3504&SA${\underline{\rm B}}$&1&${\underline{\rm r}}$s&A&AIS&A\\
NGC~3513&SB&5&r${\underline{\rm s}}$&A&GII&A\\
NGC~3547&SB&6&rs&A&GII&$-$\\
NGC~3583&SAB&3&rs&A&AIS&$-$\\
NGC~3611&SA&1&r&A&MIS&$-$\\
NGC~3614&SA&4&r&A&AIS&$-$\\
NGC~3637&SB&-1&rl&P?&AIS&$-$\\
NGC~3642&SA&2&rl&P?&GII&P?\\
NGC~3664&SB&9&r${\underline{\rm s}}$&A&MIS&$-$\\
NGC~3673&SA${\underline{\rm B}}$&1&rs&A&AIS&$-$\\
NGC~3681&SAB&1&rs&A&AIS&$-$\\
NGC~3683A&SAB&4&rs&A&AIS&$-$\\
NGC~3687&SA${\underline{\rm B}}$&2&rs&A&AIS&$-$\\
NGC~3691&SB&9&r&A&AIS&$-$\\
NGC~3705&SAB&3&${\underline{\rm r}}$s&A&AIS&A\\
NGC~3726&SA${\underline{\rm B}}$&4&r&A&AIS&A\\
NGC~3729&SB&0&r&A&GII&A\\
NGC~3780&SA&4&rs&A&AIS&$-$\\
NGC~3782&SB&8&rs&A&AIS&A\\
NGC~3786&SA&0&r&A&AIS&$-$\\
NGC~3870&SB&-2&rs&A?&GII&$-$\\
NGC~3887&SA${\underline{\rm B}}$&4&rs&A&GII&A\\
NGC~3888&SA&3&rs&A&AIS&$-$\\
NGC~3892&SA${\underline{\rm B}}$&-1&r$^{\prime}$l&P&AIS&$-$\\
NGC~3900&SA&0&r&A&AIS&$-$\\
NGC~3945&SB&-1&rl&A&AIS&$-$\\
NGC~3949&SAB&5&r${\underline{\rm s}}$&A&AIS&A\\
NGC~4030&SA&4&rs&A&MIS&A\\
NGC~4037&SAB&5&rs&A&AIS&A\\
NGC~4041&S${\underline{\rm A}}$B&5&rs&A&GII&A\\
NGC~4045&SAB&2&${\underline{\rm r}}$s&A&GII&A\\
NGC~4050&SA${\underline{\rm B}}$&1&${\underline{\rm r}}$s&A&AIS&$-$\\
NGC~4051&SAB&3&rs&A&AIS&A\\
NGC~4067&SB&2&rs&A&GII&$-$\\
NGC~4116&SB&7&rs&A&MIS&A\\
NGC~4123&SB&3&rs&A&AIS&A\\
NGC~4136&S${\underline{\rm A}}$B&4&rs&A&GII&A\\
NGC~4138&SA&-1&r&A&NGS&$-$\\
NGC~4141&SB&7&r${\underline{\rm s}}$&A&AIS&$-$\\
NGC~4145&SAB&7&rs&A&AIS&A\\
NGC~4162&SA&5&r&A?&AIS&$-$\\
NGC~4189&SA${\underline{\rm B}}$&4&r${\underline{\rm s}}$&A&GII&A\\
NGC~4212&SA&3&rs&A&GII&A\\
NGC~4234&SB&9&rs&A&AIS&A\\
NGC~4245&SB&-1&r&A&GII&A\\
NGC~4250&SAB&-1&rl&A&AIS&$-$\\
NGC~4298&SA&4&rs&A&GII&A\\
NGC~4303&SAB&5&rs&A&NGS&A\\
NGC~4309&SAB&-2&rl&A&GII&$-$\\
NGC~4314&SB&1&rl&P&NGS&P\\
NGC~4321&SAB&4&r${\underline{\rm s}}$&A&GII&A\\
NGC~4336&SAB&0&r&P&AIS&$-$\\
NGC~4339&SA&-2&r&A?&AIS&$-$\\
NGC~4340&SB&-1&r&P?&AIS&$-$\\
NGC~4371&SB&-2&r&P&DIS&$-$\\
NGC~4380&SA&2&r&A&GII&A\\
NGC~4385&SA${\underline{\rm B}}$&2&rs&A&AIS&A\\
NGC~4394&SB&0&${\underline{\rm r}}$s&A&AIS&A\\
NGC~4405&SA${\underline{\rm B}}$&1&${\underline{\rm r}}$s&A&NGS&A\\
NGC~4411A&SB&6&rs&A&GII&$-$\\
NGC~4412&SA${\underline{\rm B}}$&4&r${\underline{\rm s}}$&A&AIS&A\\
NGC~4413&SB&2&${\underline{\rm r}}$s&A&NGS&A\\
NGC~4414&SA&4&rl&A&NGS&A\\
NGC~4416&SB&8&rs&A&GII&A\\
NGC~4430&SAB&8&${\underline{\rm r}}$s&A&AIS&$-$\\
NGC~4450&SAB&1&rs&A&AIS&A\\
NGC~4454&SAB&0&r&A&AIS&$-$\\
NGC~4477&SB&1&r&P?&GII&$-$\\
NGC~4491&SB&0&rs&P?&NGS&$-$\\
NGC~4492&SA&-3&${\underline{\rm r}}$s&A&AIS&A\\
NGC~4496A&SB&7&r${\underline{\rm s}}$&A&AIS&$-$\\
NGC~4498&SB&7&rs&A&GII&A\\
NGC~4501&SA&3&r${\underline{\rm s}}$&A&AIS&A\\
NGC~4504&SAB&5&rs&A&GII&$-$\\
NGC~4519&SAB&6&r${\underline{\rm s}}$&A&GII&A\\
NGC~4528&SB&-2&r&P?&NGS&$-$\\
NGC~4531&SA&1&${\underline{\rm r}}$s&A&NGS&A\\
NGC~4540&SAB&9&rs&A&AIS&A\\
NGC~4548&SB&1&rs&A&GII&A\\
NGC~4567&SA&4&rs&A&GII&A\\
NGC~4579&SB&1&r${\underline{\rm s}}$&A&NGS&A\\
NGC~4580&SA&1&${\underline{\rm r}}$s,rs&P,A&GII&P,A\\
NGC~4593&SB&1&rs&A&AIS&$-$\\
NGC~4596&SB&0&rs&P&GII&$-$\\
NGC~4618&SB&9&r${\underline{\rm s}}$&A&NGS&A\\
NGC~4639&SB&2&${\underline{\rm r}}$s&A&AIS&A\\
NGC~4643&SB&-2&r&P?&AIS&P\\
NGC~4651&SA&4&${\underline{\rm r}}$s&A&GII&A\\
NGC~4654&SB&6&r${\underline{\rm s}}$&A&AIS&A\\
NGC~4680&SAB&3&rs&A&AIS&$-$\\
NGC~4698&SA&0&r,r&A,P&GII&A,P\\
NGC~4713&SA${\underline{\rm B}}$&5&rs&A&AIS&A\\
NGC~4725&SAB&1&r&A&AIS&A\\
NGC~4736&S${\underline{\rm A}}$B&1&rl&A&NGS&A\\
NGC~4750&SA&1&${\underline{\rm r}}$s&A&AIS&A\\
NGC~4772&SA&0&r&A&MIS&A\\
NGC~4779&SB&3&rs&A&AIS&$-$\\
NGC~4793&SA&5&rs&A&GII&A\\
NGC~4800&SA&1&rs&A&AIS&A\\
NGC~4814&SA&4&r${\underline{\rm s}}$&A&GII&$-$\\
NGC~4826&SA&1&rs,r&P,A&CAI&P,A\\
NGC~4880&S${\underline{\rm A}}$B&-1&rl&P&AIS&$-$\\
NGC~4897&SA${\underline{\rm B}}$&3&${\underline{\rm r}}$s&A&GII&$-$\\
NGC~4902&SB&3&${\underline{\rm r}}$s&A&AIS&$-$\\
NGC~4941&SA&0&${\underline{\rm r}}$s&A&AIS&$-$\\
NGC~4961&SB&4&rs&A&GII&$-$\\
NGC~4995&SAB&2&rs&A&AIS&$-$\\
NGC~5033&SA&5&rs&A&AIS&A\\
NGC~5055&SA&4&rs,rl&A,A&NGS&A,A\\
NGC~5068&SB&7&r${\underline{\rm s}}$&A&GII&A\\
NGC~5101&SB&0&${\underline{\rm r}}$s&A&AIS&$-$\\
NGC~5105&SAB&6&r${\underline{\rm s}}$&A&AIS&$-$\\
NGC~5112&SB&6&r${\underline{\rm s}}$&A&AIS&A\\
NGC~5134&SA${\underline{\rm B}}$&0&rs&A&AIS&$-$\\
NGC~5145&SA&-1&r&A&AIS&$-$\\
NGC~5194&S${\underline{\rm A}}$B&4&rs&A&GII&A\\
NGC~5195&SAB&0&r&P?&GII&P?\\
NGC~5205&SB&2&rs&A&AIS&$-$\\
NGC~5218&SB&1&rs&A&AIS&$-$\\
NGC~5300&S${\underline{\rm A}}$B&5&r${\underline{\rm s}}$&A&AIS&$-$\\
NGC~5313&SA&3&r&A&AIS&$-$\\
NGC~5334&SB&6&rs&A&AIS&A\\
NGC~5339&SB&2&rs&A&AIS&$-$\\
NGC~5347&SB&1&${\underline{\rm r}}$s&A&AIS&A\\
NGC~5350&SB&3&${\underline{\rm r}}$s&A&AIS&$-$\\
NGC~5364&SA&3&r&A&DIS&A\\
NGC~5371&SAB&3&r${\underline{\rm s}}$&A&AIS&A\\
NGC~5375&SB&1&${\underline{\rm r}}$s&A&GII&$-$\\
NGC~5376&SA&2&${\underline{\rm r}}$s&A&AIS&$-$\\
NGC~5383&SB&1&rs&A&AIS&$-$\\
NGC~5426&S${\underline{\rm A}}$B&5&rs&A&GII&$-$\\
NGC~5457&S${\underline{\rm A}}$B&5&r${\underline{\rm s}}$&A&GII&A\\
NGC~5534&SB&1&r${\underline{\rm s}}$&A&AIS&A\\
NGC~5595&SAB&6&r${\underline{\rm s}}$&A&AIS&$-$\\
NGC~5600&SB&8&rs&A&AIS&$-$\\
NGC~5636&SAB&0&r&A&MIS&$-$\\
NGC~5665&SAB&5&rs&A&AIS&$-$\\
NGC~5668&SAB&6&rs&A&CAI&A\\
NGC~5669&SB&7&rs&A&AIS&A\\
NGC~5678&SA&3&${\underline{\rm r}}$s&A&AIS&$-$\\
NGC~5701&SB&0&rl&P&MIS&$-$\\
NGC~5713&SB&2&rs&A&MIS&A\\
NGC~5728&SB&0&${\underline{\rm r}}$s&A&AIS&A\\
NGC~5740&SA${\underline{\rm B}}$&2&r&A&MIS&$-$\\
NGC~5757&SB&2&rs&A&AIS&A\\
NGC~5768&SAB&4&rs&A&MIS&$-$\\
NGC~5770&SAB&-1&r&P&GII&$-$\\
NGC~5806&SAB&2&rs&A&GII&A\\
NGC~5821&S${\underline{\rm A}}$B&1&r&A&AIS&$-$\\
NGC~5850&SB&2&r&A&MIS&A\\
NGC~5892&SB&6&r${\underline{\rm s}}$&A&AIS&$-$\\
NGC~5915&SA&5&r&A&GII&A\\
NGC~5921&SB&3&${\underline{\rm r}}$s&A&AIS&A\\
NGC~5930&SAB&0&rs&A&AIS&$-$\\
NGC~5957&SB&1&${\underline{\rm r}}$s&A&GII&$-$\\
NGC~5962&SAB&5&rs&A&NGS&A\\
NGC~5964&SB&6&r${\underline{\rm s}}$&A&GII&A\\
NGC~6012&SB&2&r&A&GII&$-$\\
NGC~6014&SA${\underline{\rm B}}$&0&rs&A?&AIS&$-$\\
NGC~6267&SB&3&rs&A&AIS&$-$\\
NGC~6278&SA&-2&r&P?&AIS&$-$\\
NGC~6412&SB&6&rs&A&AIS&A\\
NGC~6902&S${\underline{\rm A}}$B&1&${\underline{\rm r}}$s&A&NGS&$-$\\
NGC~7070&SB&5&r${\underline{\rm s}}$&A&AIS&$-$\\
NGC~7098&SAB&0&r$^{\prime}$l&A&AIS&A\\
NGC~7107&SB&8&r${\underline{\rm s}}$&A&AIS&$-$\\
NGC~7179&SB&0&r&A&GII&$-$\\
NGC~7205&SA&4&rs&A&AIS&A\\
NGC~7290&SA&3&rs&A&AIS&$-$\\
NGC~7378&SA&0&${\underline{\rm r}}$s&A&AIS&$-$\\
NGC~7418&SAB&5&rs&A&NGS&$-$\\
NGC~7421&SB&2&rs&A&NGS&$-$\\
NGC~7424&SB&6&r${\underline{\rm s}}$&A&GII&A\\
NGC~7496&SB&3&r${\underline{\rm s}}$&A&NGS&$-$\\
NGC~7513&SB&1&rs&A&AIS&$-$\\
NGC~7531&SAB&1&r&A&GII&$-$\\
NGC~7552&SB&1&r${\underline{\rm s}}$&A&NGS&A\\
NGC~7716&S${\underline{\rm A}}$B&2&r&A&MIS&$-$\\
NGC~7723&SB&2&rs&A&MIS&$-$\\
NGC~7742&SA&-1&r,r&A,A&GII&$-$\\
NGC~7743&SAB&1&${\underline{\rm r}}$s&P&AIS&$-$\\
NGC~7755&SA${\underline{\rm B}}$&4&rs&A&AIS&$-$\\
PGC~003853&SB&6&${\underline{\rm r}}$s&A&AIS&$-$\\
PGC~006626&SB&6&rs&A&AIS&$-$\\
PGC~012633&S${\underline{\rm A}}$B&2&rs&A&MIS&$-$\\
PGC~012664&SA${\underline{\rm B}}$&6&r${\underline{\rm s}}$&A&MIS&$-$\\
PGC~032091&S${\underline{\rm A}}$B&5&r${\underline{\rm s}}$&A&AIS&$-$\\
PGC~038250&SAB&9&rs&A&AIS&$-$\\
PGC~044735&SAB&8&rs&A&GII&$-$\\
PGC~044952&SA&7&r&A&AIS&$-$\\
PGC~047721&SA&2&${\underline{\rm r}}$s,r,r&A,A?,A?&AIS&$-$\\
PGC~048179&SB&6&r${\underline{\rm s}}$&A&GII&A\\
PGC~054944&SB&7&${\underline{\rm r}}$s&A&AIS&$-$\\
UGC~01551&SB&6&rs&A&GII&$-$\\
UGC~04867&SB&7&r${\underline{\rm s}}$&A&AIS&$-$\\
UGC~06023&SAB&5&${\underline{\rm r}}$s&A&AIS&A\\
UGC~06309&SB&5&r${\underline{\rm s}}$&A&AIS&$-$\\
UGC~07184&SB&7&rs&A&MIS&$-$\\
UGC~08155&SA&1&rs&A&AIS&$-$\\
UGC~09356&SAB&4&rs&A&AIS&$-$\\
UGC~09569&SB&5&r${\underline{\rm s}}$&A&AIS&$-$\\
UGC~10054&SB&7&rs&A&AIS&A\\
UGC~10791&SB&7&rs&A&NGS&$-$\\
UGC~12151&SB&7&r${\underline{\rm s}}$&A&MIS&$-$\\